\documentclass[twocolumn,showpacs,prl,nofootinbib,superscriptaddress,amsmath,amssymb,floatfix]{revtex4}

\usepackage{graphicx}  % needed for figures
\usepackage{dcolumn}   % needed for some tables
\usepackage{bm}        % for math
\usepackage{amssymb}   % for math

\newcommand{\BibitemShut}[1]{}

\def \aap  {A\&A}

\def \apj  {ApJ}
\def \apjs  {ApJS}
\def \apjl  {ApJL}

\begin{document}

% The following information is for internal review, please remove them for submission
\widetext
\leftline{Version 00 as of \today}
\leftline{To be submitted to PRL}
%\leftline{Comment to {\tt d0-run2eb-nnn@fnal.gov} by xxx, yyy}

% the following line is for submission, including submission to the arXiv!!
%\hspace{5.2in} \mbox{Fermilab-Pub-04/xxx-E}

\title{Resolving the Extragalactic $\gamma$-ray Background above 50\,GeV with
{\it Fermi}-LAT}
%\input author_list.tex       % D0 authors (remove the first 3 lines
                             % of this file prior to submission, they
                             % contain a time stamp for the authorlist)
                             % (includes institutions and visitors)

\author{M.~Ackermann}
\affiliation{Deutsches Elektronen Synchrotron DESY, D-15738 Zeuthen, Germany}
\author{M.~Ajello}
\email{majello@slac.stanford.edu}
\affiliation{Department of Physics and Astronomy, Clemson University, Kinard Lab of Physics, Clemson, SC 29634-0978, USA}
\author{A.~Albert}
\affiliation{W. W. Hansen Experimental Physics Laboratory, Kavli Institute for Particle Astrophysics and Cosmology, Department of Physics and SLAC National Accelerator Laboratory, Stanford University, Stanford, CA 94305, USA}
\author{W.~B.~Atwood}
\affiliation{Santa Cruz Institute for Particle Physics, Department of Physics and Department of Astronomy and Astrophysics, University of California at Santa Cruz, Santa Cruz, CA 95064, USA}
\author{L.~Baldini}
\affiliation{Universit\`a di Pisa and Istituto Nazionale di Fisica Nucleare, Sezione di Pisa I-56127 Pisa, Italy}
\affiliation{W. W. Hansen Experimental Physics Laboratory, Kavli Institute for Particle Astrophysics and Cosmology, Department of Physics and SLAC National Accelerator Laboratory, Stanford University, Stanford, CA 94305, USA}
\author{J.~Ballet}
\affiliation{Laboratoire AIM, CEA-IRFU/CNRS/Universit\'e Paris Diderot, Service d'Astrophysique, CEA Saclay, F-91191 Gif sur Yvette, France}
\author{G.~Barbiellini}
\affiliation{Istituto Nazionale di Fisica Nucleare, Sezione di Trieste, I-34127 Trieste, Italy}
\affiliation{Dipartimento di Fisica, Universit\`a di Trieste, I-34127 Trieste, Italy}
\author{D.~Bastieri}
\affiliation{Istituto Nazionale di Fisica Nucleare, Sezione di Padova, I-35131 Padova, Italy}
\affiliation{Dipartimento di Fisica e Astronomia ``G. Galilei'', Universit\`a di Padova, I-35131 Padova, Italy}
\author{K.~Bechtol}
\affiliation{Dept.  of  Physics  and  Wisconsin  IceCube  Particle  Astrophysics  Center, University  of  Wisconsin, Madison,  WI  53706, USA}
\author{R.~Bellazzini}
\affiliation{Istituto Nazionale di Fisica Nucleare, Sezione di Pisa, I-56127 Pisa, Italy}
% ...

\date{\today}

\begin{abstract}
%The {\it Fermi} Large Area Telescope (LAT) Collaboration has recently released a catalog of 360 sources, at high-latitude mostly blazars, detected above 50 GeV (2FHL). The 2FHL has been obtained using 80\,months of data re-processed with Pass~8, the newest event-level analysis, which significantly improves the acceptance and angular resolution of the instrument.
The {\it Fermi} Large Area Telescope (LAT) Collaboration has recently released a catalog of 360 sources detected above 50 GeV (2FHL). This catalog was obtained using 80 months of data re-processed with Pass~8, the newest event-level analysis, which significantly improves the acceptance and angular resolution of the instrument. Most of the 2FHL sources at high Galactic latitude are blazars.
Using detailed Monte Carlo simulations, we measure, for the first time, the source count distribution, $dN/dS$, of extragalactic $\gamma$-ray sources at $E>50$ GeV and find that it is compatible with a Euclidean distribution down to the lowest measured source flux in the 2FHL ($\sim8\times 10^{-12}$\,ph cm$^{-2}$ s$^{-1}$). We employ a one-point photon fluctuation analysis to constrain the behavior of $dN/dS$ below the source detection threshold. Overall the source count distribution is constrained over three decades in flux and found compatible with a broken power law with a break flux, $S_b$, in the range $[8 \times 10^{-12},1.5 \times 10^{-11}]$ ph cm$^{-2}$ s$^{-1}$ and power-law indices below and above the break of $\alpha_2 \in [1.60,1.75]$ and $\alpha_1 = 2.49 \pm 0.12$ respectively.
Integration of $dN/dS$ shows that point sources account 
for at least $86^{+16}_{-14}\%$ of the total extragalactic $\gamma$-ray background.
The simple form of the derived source count distribution is consistent with a single population
(i.e. blazars) dominating the source counts to the minimum flux explored by this analysis. We estimate the density of sources detectable in blind surveys that will be performed in the coming years by the Cherenkov Telescope Array.
\end{abstract}

\maketitle

%\section{Introduction}
% sections are not used for PRL papers

The origin of the extragalactic $\gamma$-ray background (EGB),  the
Universe's glow in $\gamma$ rays, has been debated since the first measurement with the SAS-2 satellite \cite{fichtel95}. The EGB spectrum has been
accurately measured, from 100\,MeV to 820\,GeV, by
the Large Area Telescope (LAT) on board the {\it Fermi} Gamma-Ray Space Telescope mission \cite{Ackermann:2014usa}.
Part of the EGB arises from the emission of resolved and unresolved point sources
like blazars, star-forming and radio galaxies \citep[e.g.][]{dermer07,Ajello:2015mfa,DiMauro:2015tfa}, which are routinely detected in $\gamma$ rays. A possible contribution to the EGB may also come from diffuse processes such as annihilating/decaying dark matter particles (see \cite{Fornasa:2015qua} for a review).
%Mattia commented the following two sentences
%Part of the EGB is thought to arise from the emission of unresolved  point sources like blazars, star forming and radio galaxies \citep[e.g.][]{dermer07,Ajello:2015mfa,DiMauro:2015tfa}, which are routinely detected at $\gamma$ rays.
%The EGB may also comprise many other contributions like the emission from intergalactic shocks \citep{loeb00,miniati02}, the cascade $\gamma$-ray emission originated from the interaction of ultra high-energy cosmic rays (UHECRs) with the cosmic microwave background \citep{bhattacharjee00}, and e.g. the emission from dark matter annihilation \citep{ullio02}. 
%Unveiling the origin of the EGB is a challenging task because no previous instrument or analysis has been sensitive enough to resolve the majority of the EGB.

\begin{figure*}
	\centering
\includegraphics[width=1.03\columnwidth]{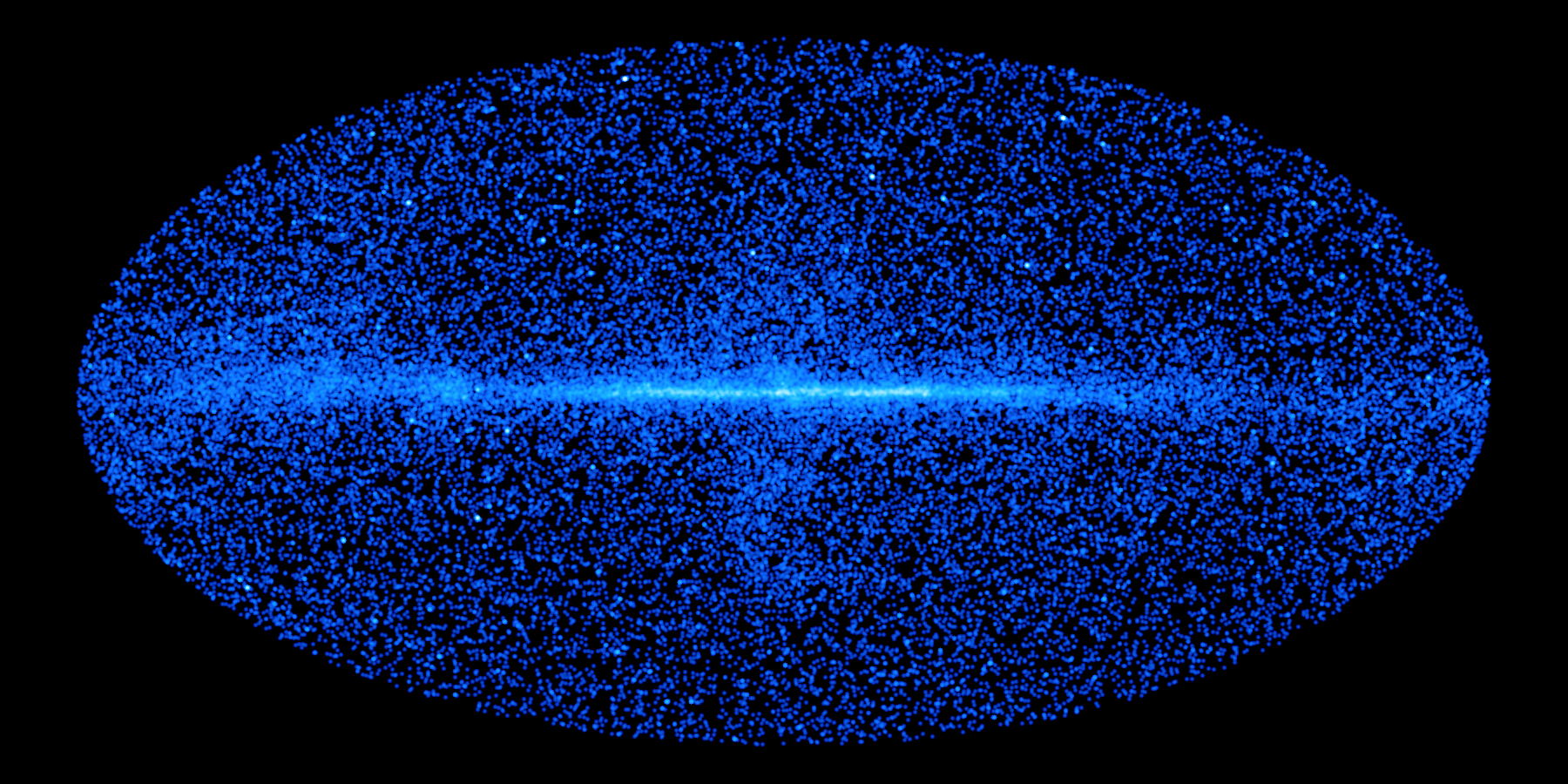}
\includegraphics[width=1.03\columnwidth]{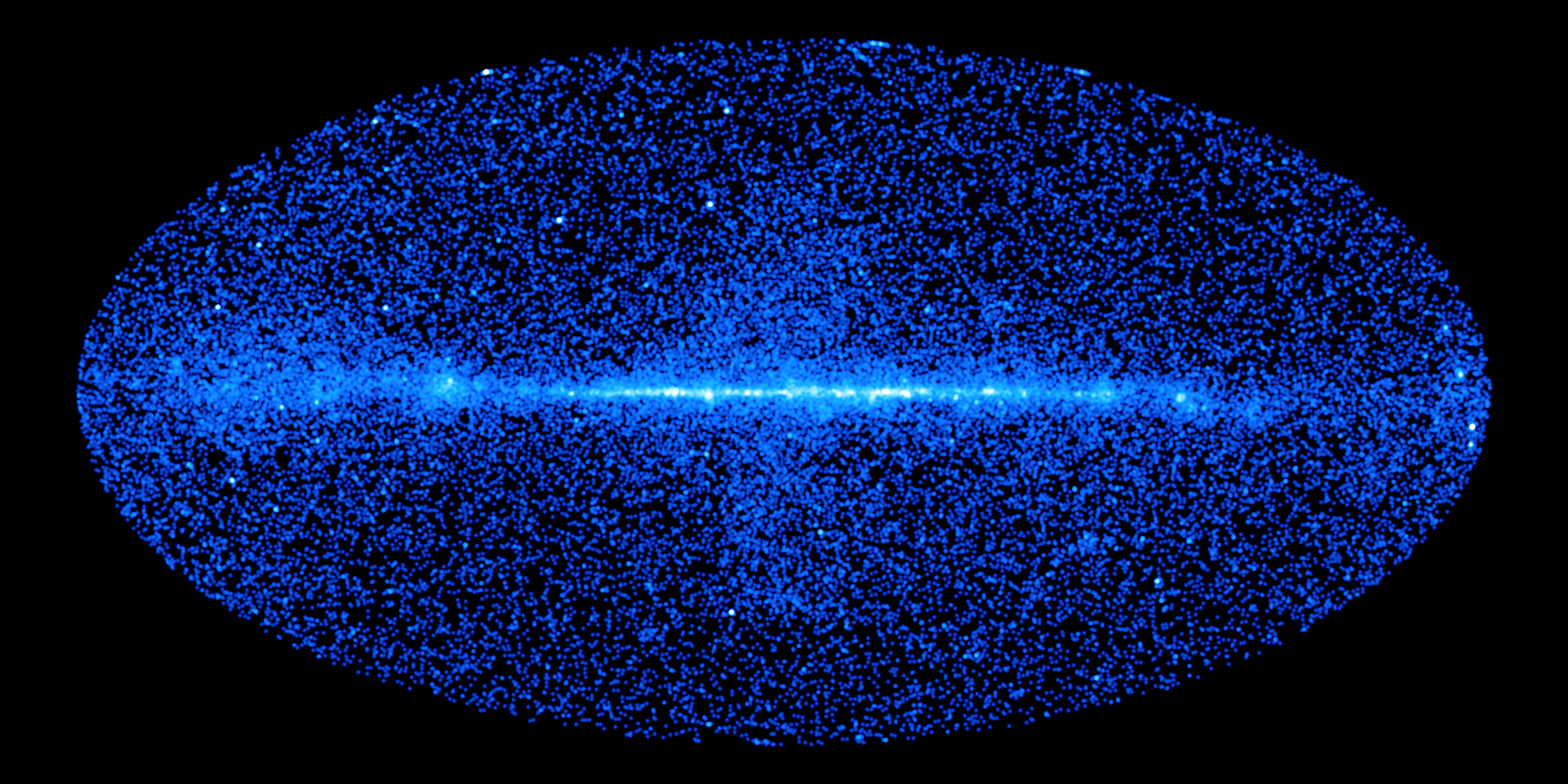}
\caption{In the left (right) panel the adaptively smoothed count map of one simulation (real sky) in the energy range 50 GeV-2 TeV is represented in Galactic coordinates and Hammer-Aitoff projection. The two maps contain about 60000 $\gamma$-ray events.}
\label{fig:map} 
\end{figure*}

Here we show for the first time that {\it Fermi}-LAT is able to resolve the high-energy EGB into point-like sources. 
Indeed, thanks to the accrual of 80\,months of data (see right panel of Fig.~\ref{fig:map}) and the increased acceptance and improved point-spread function delivered by the new event-level analysis dubbed Pass~8 \cite{Atwood:2013dra}, the LAT has recently performed an all-sky survey at $>$50\,GeV resulting in the detection of 360 $\gamma$-ray sources that constitute
the second catalog of hard {\it Fermi}-LAT sources \citep[2FHL,][]{2fhl}

Blazars, mostly belonging to the BL Lacertae (BL Lac) population, are the majority (74\,$\%$) of the sources in the 2FHL catalog. At Galactic latitudes ($b$) larger than 10$^{\circ}$ about 70$\%$ of the detected sources are associated with BL Lacs. Only 7\,\% of the high-latitude ($|b|>10^{\circ}$) sources are classified as something other than
BL Lacs, 4\% of them being Flat Spectrum Radio Quasars (FSRQs),  while blazars of uncertain type and unassociated sources constitute the remaining 23\,\% of the sample. The median of the synchrotron peak frequencies for blazars of uncertain type is very similar to that of BL Lacs (log$_{\rm 10}(\nu^{S}_{peak}/{\rm Hz})=15.7$ vs. 15.6). 
The same holds for the median spectral index of unassociated sources ($\Gamma=$3.0 vs. 3.1).
%We thus believe that the majority of blazars of uncertain type and unassociated sources are BL Lacs.
This is supporting the fact that blazars of uncertain type and unassociated sources are almost entirely BL Lacs.
Therefore, the fraction of likely blazars in the high-latitude 2FHL sample is 97\,\% (93\% BL Lacs and 4\% FSRQs).

%INTRODUCTION
%In this paper we derive, for the first time, the source count distribution $dN/dS$ of sources located at latitude $|b|>10^{\circ}$, where $S$ is the photon flux (in unit of ph\,cm$^{-2}$\,s$^{-1}$) measured in the 50\,GeV--2\,TeV energy band.

In this paper, we derive the source detection efficiency of the 2FHL catalog analysis using accurate Monte Carlo simulations of the $\gamma$-ray sky. We then infer the intrinsic flux distribution $dN/dS$ of sources located at a latitude $|b| > 10^{\circ}$, where $S$ is the photon flux (ph\,cm$^{-2}$\,s$^{-1}$) measured in the 50\,GeV--2\,TeV energy band.

%In this paper we derive, for the first time, the detection efficiency, determined via accurate Monte Carlo simulations that resemble as closely as possible the real sky, and consequently we infer the source count distribution $dN/dS$ of sources located at a latitude $|b|>10^{\circ}$, where $S$ is the photon flux (in unit of ph\,cm$^{-2}$\,s$^{-1}$) measured in the 50\,GeV--2\,TeV energy band.

%The source count distribution is of interest to assess the contribution of sources to the {\bf EGB} %\citep{Ackermann:2014usa}, the number of sources detectable in blind surveys by future experiments like the Cherenkov Telescope Array (CTA) \cite{2013APh....43..317D}, and e.g. the angular anisotropies of the VHE sky.

\begin{figure}[t]
	\centering
	\includegraphics[width=1.03\columnwidth]{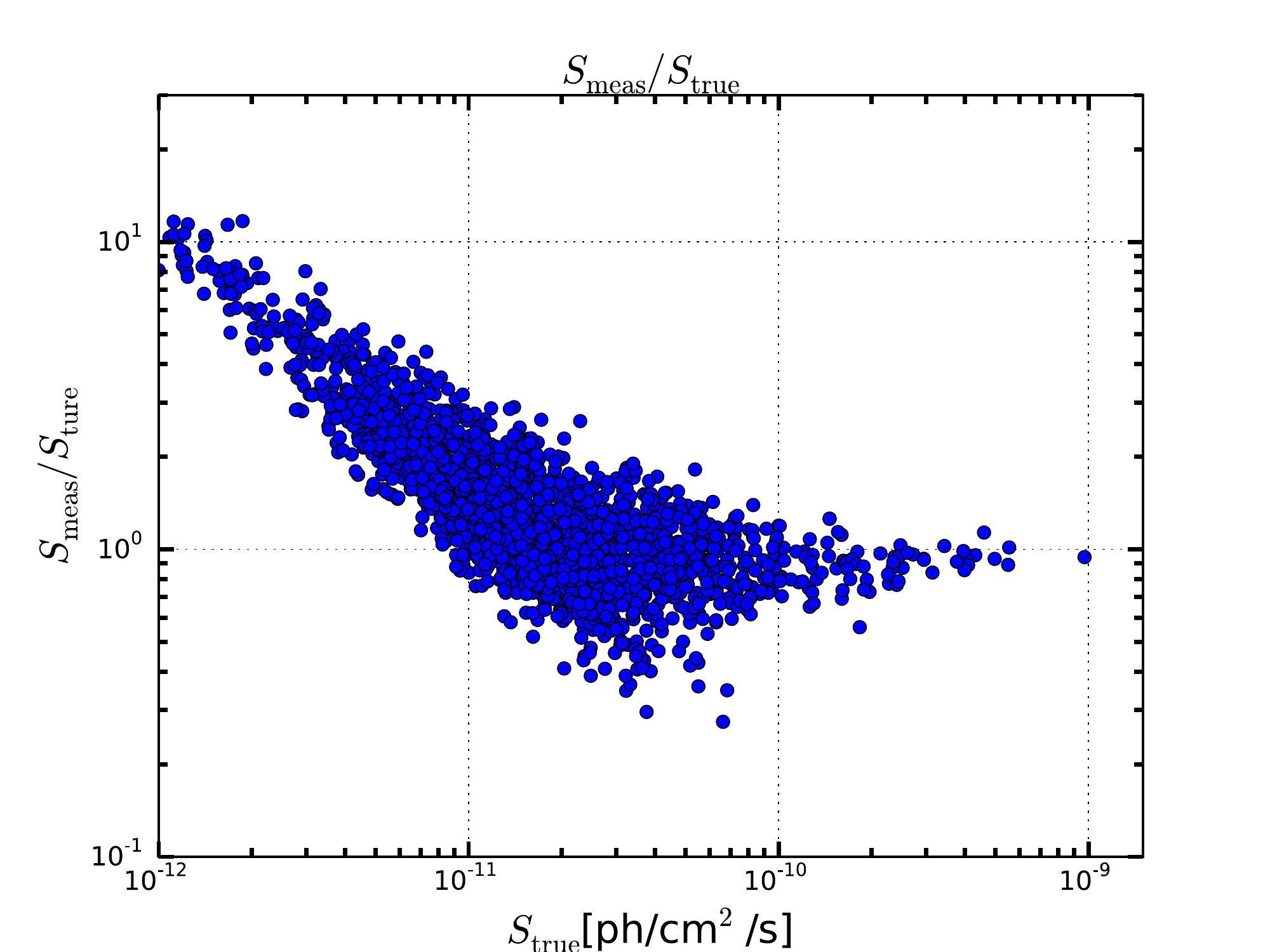}
	\includegraphics[width=1.03\columnwidth]{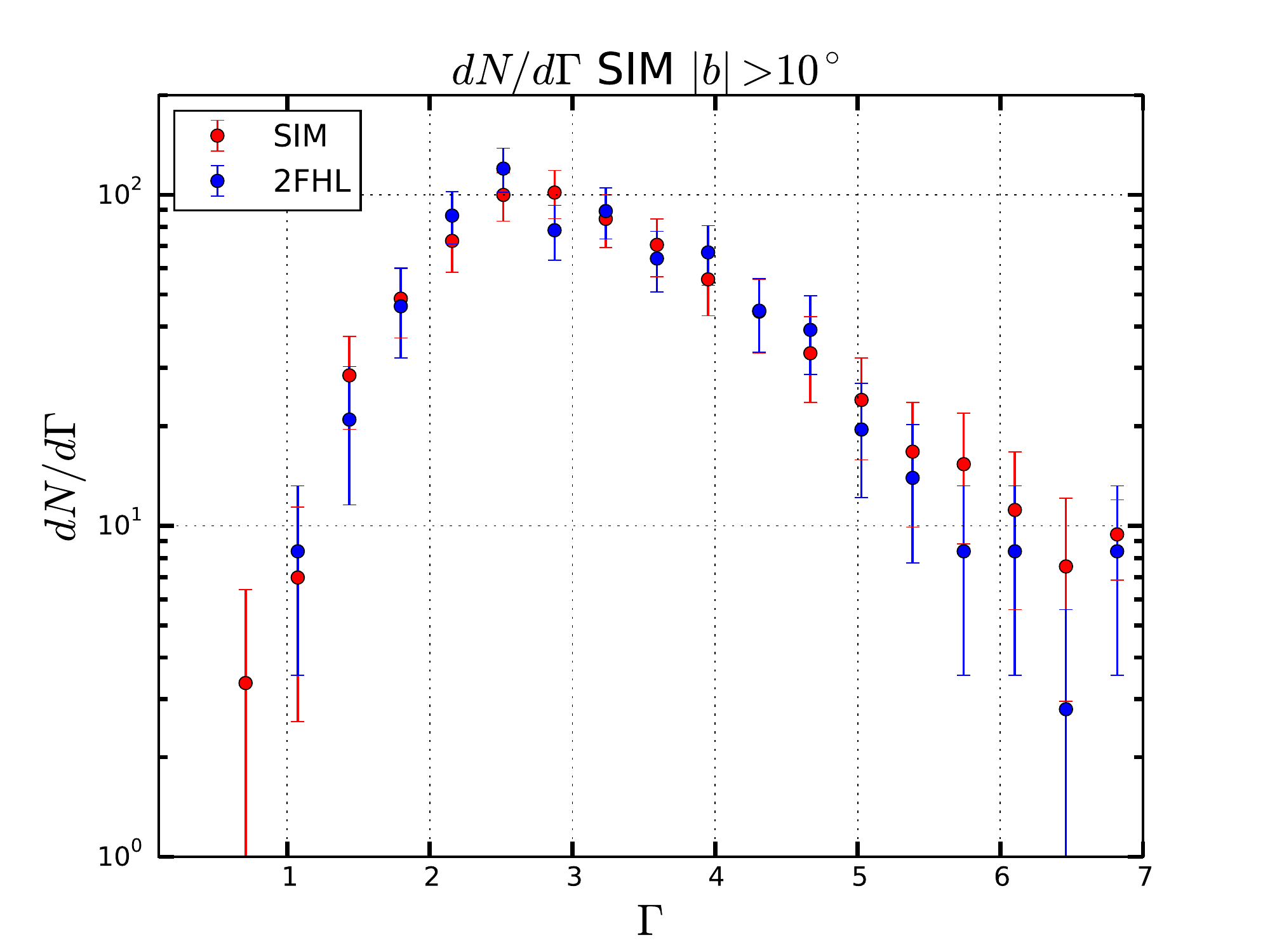}
\caption{Left Panel: ratio of the measured-to-simulated source flux (as derived from the analysis of the simulations described in the text) as a function of simulated source flux.
%In the left panel we show the ratio of the source flux derived with the analysis  $S_{\rm{out}}$ and the intrinsic source flux of the simulations $S_{\rm{sim}}$. 
Right Panel: comparison between the photon index distributions of sources detected in 2FHL (blue points) and the average of the simulations (red points).}
\label{fig:checks} 
\end{figure}

%SIMULATIONS
%Measuring the source count distribution requires knowledge of the {\bf detection efficie ncy}, which is best determined via accurate Monte Carlo simulations. 
%In this work we perform and analyze Monte Carlo simulations of the $>$50\,GeV sky that resemble as closely as possible the real sky. 
The simulations were performed using the {\tt gtobssim} tool, which is part of the {\it Fermi} ScienceTools distribution, and using the same pointing and live time history and event selection as used in the 2FHL catalog.
%In the first part of the analysis we use the {\it Fermi} Science Tool {\tt GTOBSSIM} to create realistic simulations of the $\gamma$-ray sky.
We have employed the P8R2\_SOURCE\_V6 instrument response function for the simulations and analysis and the Galactic and isotropic diffuse emission were simulated using the
 {\tt gll\_iem\_v06.fits} and {\tt iso\_P8R2\_SOURCE\_V6\_v06.txt} templates
 \footnote{See http://fermi.gsfc.nasa.gov/ssc/}.
The last ingredient of the simulations is an isotropic population of point sources that has the characteristics of blazars (fluxes and spectra) as detected in 2FHL. 
%Mattia has commented the following sentence
%This requires knowledge of the source count distribution itself, which is the main goal of this analysis and was not known a-priori, so the process required iterating till the source count distribution was characterized and could be inputed in our final run of Monte Carlo simulations. 
%The simulations described here rely on the source count distribution as determined at the end of this analysis (see later), which is, at fluxes $\gtrsim 10^{-11}$ ph cm$^{-2}$ s$^{-1}$, approximately Euclidean:  namely a power-law with a slope $\alpha_1=5/2$ where $dN/dS \propto S^{-\alpha_1}$. 
The simulations described here were produced iteratively and ultimately rely on the source count distribution $dN/dS \propto S^{-\alpha}$ as determined at the end of photon fluctuation analysis (see later), which is, a broken power law with a break flux $S_b =1\times 10^{-11}$ ph cm$^{-2}$ s$^{-1}$ and a Euclidean slope above the break, $\alpha_1=5/2$, while below $S_b$ the slope is $\alpha_2=1.65$.
%we approximate above the  sensitivity flux of the 2FHL catalog ($\approx 10^{-11}$ ph cm$^{-2}$ s$^{-1}$) the source count distribution with a Euclidean distribution:
%but we have first to derive their flux distribution. We consider above the sensitivity flux of the 2FHL ($\approx 10^{-11}$ ph cm$^{-2}$ s$^{-1}$) the same shape as in the catalog,
% namely a power-law with a slope $\alpha_1=$2.50 where $dN/dS \propto S^{-\alpha_1}$. 
Sources were generated with fluxes in the range $[S_{\rm{min}},S_{\rm{max}}]=[10^{-14},10^{-9}]$ ph cm$^{-2}$ s$^{-1}$ and with power-law spectra of the form $dN/dE\propto E^{-\Gamma}$.
%We decided to start creating sources from $1.5 \times 10^{-12}$ ph cm$^{-2}$ s$^{-1}$ because with a slope value of $-2.5$ and integrating the differential flux distribution from this flux value, $\int_{S_{\rm{min}}} dN/dS dS$, we generate about the $100\%$ of the Extragalactic $\gamma$-Ray Background (EGB).
For each source the photon index $\Gamma$ is drawn from a Gaussian distribution with average value 3.2 and standard deviation 0.7 (this reproduces the observed distribution as shown on the right panel of Fig.~\ref{fig:checks}).
Galactic sources are not considered in the simulations since we are interested in the flux distribution of blazars at $|b|>10^{\circ}$.
We produced 10 simulations of the $\gamma$-ray sky following these prescriptions and in Fig.~\ref{fig:map} the sky map of one simulation is shown together with the real one. Clearly visible in both maps are the diffuse emission along the Galactic plane, the {\it Fermi} bubbles \cite{Fermi-LAT:2014sfa}, the emission from point sources and the isotropic diffuse emission.

%\begin{figure}
%	\centering
%\includegraphics[width=1.1\columnwidth]{/Users/mattiadimauro/Dropbox/2FHLeff/plot/dNdE_comparison_ult_10deg_25bins.pdf}
%\caption{Spectral energy distribution $dN/dE$ of the simulated GDE (blue points), isotropic diffuse (green points), sources (red points) and total $\gamma$-ray emission of the simulated map (brown points) compared with the same quantity of the real sky map (black points). We show also the difference of simulated and the real sky spectra with orange points.}
%\label{fig:comp} 
%\end{figure}

%%%% Marco commented and rephrased the following
%%%%
%In order to check that our simulations are correct representations of the real sky, we have compared the average $\gamma$-ray energy spectrum $dN/dE$ of the 10 simulations with the one of the real sky. The spectra of the GDE, isotropic diffuse, sources and the differences between the simulation and the real sky $dN/dE$ are displayed. Since the GDE template ends at about 515 GeV, as clearly visible in Fig.~\ref{fig:comp}, we have renormalized the isotropic diffuse template in order to predict the correct number of photons in this energy range. The simulations provide a good representation of the real sky indeed the residuals are at most on the 10$\%$ level if $E<$1 TeV. At energies larger than 1 TeV the residuals can be as large as about 30$\%$ with however a low statistics of events (0.5$\%$ of the total $\gamma$-ray events).

The energy spectrum of the simulations is consistent within 10\,$\%$, at all energies of interest and for photons detected at $|b|>10^{\circ}$, with that of the LAT observations. As clearly visible in Fig.~\ref{fig:map}, the spatial distribution of gamma rays of the real map is also correctly reproduced.
%The simulations reproduce the morphology of the $\gamma$-ray sky (see Fig.~\ref{fig:map}) and its energy spectrum (for photons detected at $|b|>10^{\circ}$) within 10\,$\%$ at all energies of interest for this analysis.
%The simulations are good representations of the $\gamma$-ray sky since they reproduce its morphology (see Fig.~\ref{fig:map}) and its energy spectrum (for photons detected at $|b|>10^{\circ}$) within 10\,$\%$ at all energies of interest for this analysis.
The 10 simulations are analyzed exactly as the real data were for the 2FHL catalog.
This starts from detecting source candidates using a sliding-cell algorithm and a wavelet analysis \cite{2007AIPC..921..546C} then analyzing each with the standard {\it Fermi} Science Tools, in order to derive the $\gamma$-ray properties of detectable sources (see \cite{2fhl} for more details). 
%This starts from the detection of source candidates (called seeds) using a sliding-cell algorithm and a wavelet analysis \cite{2007AIPC..921..546C} to identify excesses above the background. 
%Once we have fixed the simulations we analyze them, with the standard {\it Fermi} Science tools, in order to derive the $\gamma$-ray properties of detectable sources. In the first step of the analysis we localize the potential sources with identification of source seeds via a sliding-cell algorithm as excesses above the background, as clusters at least 3 photons and via a wavelet analysis \cite{2007AIPC..921..546C}.
%The seed list contains real sources and false detections associated with statistical fluctuation of the background, so further analysis is performed to create the final source list.
%A region of interest (ROI) is created for each seed including all the photons
%within a radius of 15$^{\circ}$. A sky model, {\bf which} includes all the potential sources
%and diffuse backgrounds, namely the Galactic diffuse model and
%the isotropic diffuse model, is fitted to the data using the unbinned maximum likelihood (ML) algorithm provided with the {\it Fermi} Science Tools (version v10r0p5).
%The position of each seed is refined after the spectral parameters of each model component have been optimized. Because the significance of a source depends both on its optimized position and the spectral parameters, this procedure is repeated three times to make sure that all the parameters have been refined successfully.
As in the 2FHL catalog, detected sources are those with a test statistic (TS)$>$25 and at least 3 associated photons predicted by the likelihood fit.
This leads to the detection, in the simulations, of $271 \pm 18$ sources at $|b|>10^{\circ}$, which is in good agreement with the 253 sources detected in the 2FHL. Moreover, the simulations show that the 2FHL catalog contains at most 1\% of false detections.

%%%%%%
%%%%%% MARCO rephrased the below paragraph above.
%%%%%%
%Given the seed list we then perform a Maximum Likelihood (ML) analysis among the seeds considering a region of interest (ROI) of 15$^{\circ}$ around them. For each ROI a sky model is generated including the GDE and the isotropic diffuse emission. Finally the sky model is fitted with an unbinned ML analysis with {\it Fermi} Science Tools version v10\_r01\_00.  This process is performed three times and for each new step the result of the previous one is considered as input in order to optimize the calculation of positions of point-like sources. Considering the method described above, we achieve the detection of on average 273 sources for $|b>10^{\circ}$, with a TS$>$25 and $N_{\rm{pred}}>3$ (against 257 detected sources in the 2FHL).  $N_{\rm{pred}}>3$ is the number of events above the background and the cut on this quantity was introduced to limit to $0.1\%$ the number of false positives in the final catalog.

%COMPARISON WITH CATALOG

In order to further validate our analysis we have performed two consistency checks on the simulations.
%The first one tests the ratio of the flux as measured with the {\it Fermi} Science tool analysis with respect to the input flux of the sources in the simulations $S_{\rm{out}}/S_{\rm{sim}}$. 
The first compares the input source fluxes $S_{\rm{true}}$ with the fluxes $S_{\rm{meas}}$ measured with the {\it Fermi} Science Tools in the simulations.
The result displayed in the left panel of Fig.~\ref{fig:checks} shows that for bright sources this ratio converges to 1 as expected in the absence of biases or errors.
On the other hand $S_{\rm{meas}}/S_{\rm{true}}$ for faint sources deviates systematically from 1. This effect is readily understood as caused by the Eddington bias, which is the statistical fluctuations of sources with a simulated flux below the threshold to a flux above the detection threshold \cite{1913MNRAS..73..359E}. 
%Mattia has commented the following sentence
%Because the number of sources below the threshold is large (due to the power-law shape of the source counts), statistical {\bf fluctuation} in the number of photons detected from a given sources leads to the detection of some of the sources that experience upward fluctuations. Fig.~\ref{fig:checks} shows that the LAT detects sources with a measured flux of 10$^{-11}$\,ph cm$^{-2}$ s$^{-1}$ while their real flux can be a factor of 10 lower.
%This effect is explained by the fact that the emission of faint sources is reconstructed with very soft photon index and in turn the derived flux is larger with respect to the real one.
Our second check compares of the average photon index
distribution ($dN/d{\Gamma}$), as derived from the simulations, with the same
distribution as derived from the 2FHL catalog. This is reported in the right panel of Fig.~\ref{fig:checks} and it shows that our description of the $\gamma$-ray sky and of the blazar population is faithful to the real one.

%We have also derived the average photon index differential distribution $dN/d{\Gamma}$ of the simulations and we have compared it with the same quantity of the 2FHL catalog in right panel of Fig.~\ref{fig:checks}. It is clear from this figure that $dN/d{\Gamma}$ of the simulations is compatible with the one of the catalog meaning that the input values for the $\Gamma$ distribution of blazars give a good representation of the detected source distribution.

%EFFICIENCY
\begin{figure}[t]
	\centering
	\includegraphics[width=1.\columnwidth]{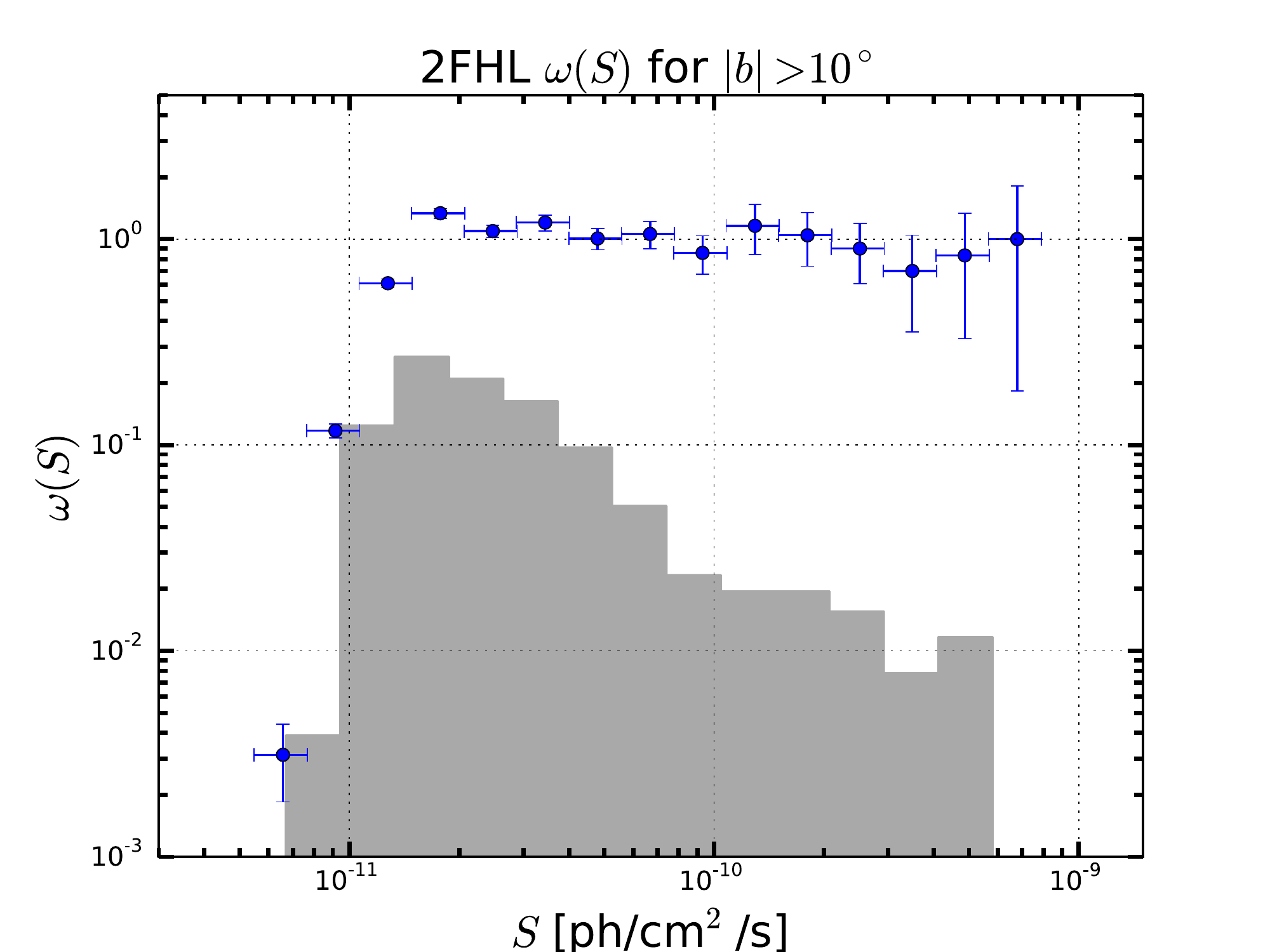}
	\includegraphics[width=1.\columnwidth]{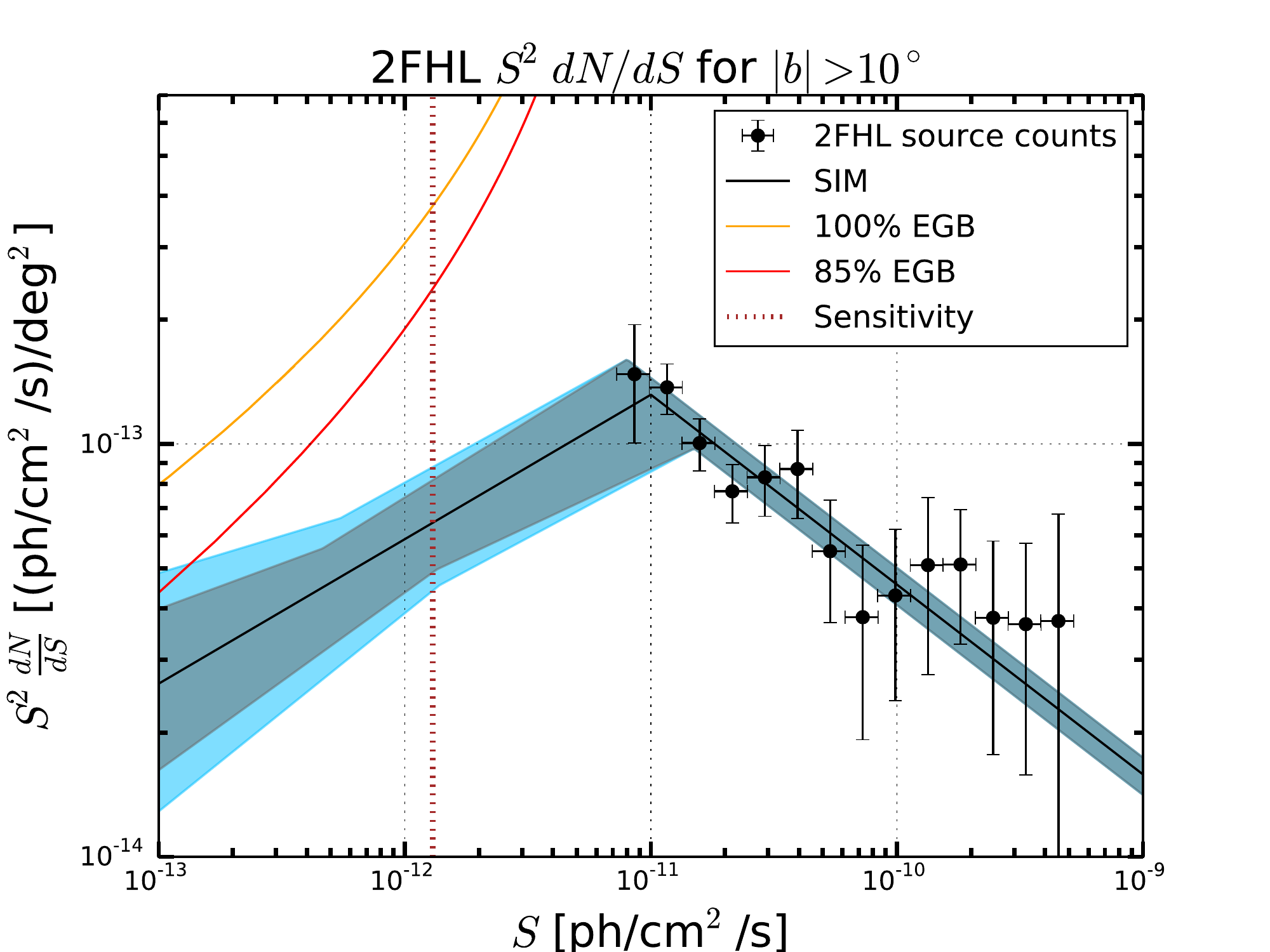}
\caption{Left Panel: detection efficiency $\omega(S)$ (blue points) as a function
of source flux and normalized distribution of source fluxes detected in 2FHL (grey shaded histogram).
Right Panel: intrinsic $S^2 dN/dS$ distribution (black points). The black solid line shows our best-fit model, while the grey and cyan bands show the $1\sigma$ and $3\sigma$ uncertainty bands from the photon fluctuation analysis. The vertical brown dotted line represents the sensitivity of the photon fluctuation analysis.
%The dashed line shows the maximum flux at which a re-steepening of the source counts to a Euclidean behavior may happen. 
The orange and red curves indicate where 85\% and 100\% of the EGB intensity above 50 GeV \cite{Ackermann:2014usa} would be produced when extrapolating the flux distribution below the break with different values of faint-end slope, $\alpha_2$.}
%In the same plot we show also our best-fit model (black line), the $1\sigma$ and $3\sigma$ uncertainty bands (grey and cyan band) derived with the pixel counting method and the upper limit to a Euclidean re-steepening of the $S^2 dN/dS$ distribution (dashed lines), which is limited to be at fluxes $\geq7\times10^{-13}$ ph/cm$^2$/s.}
\label{fig:efficiency} 
\end{figure}

\begin{figure}[t]
	\centering
	\includegraphics[width=1.\columnwidth]{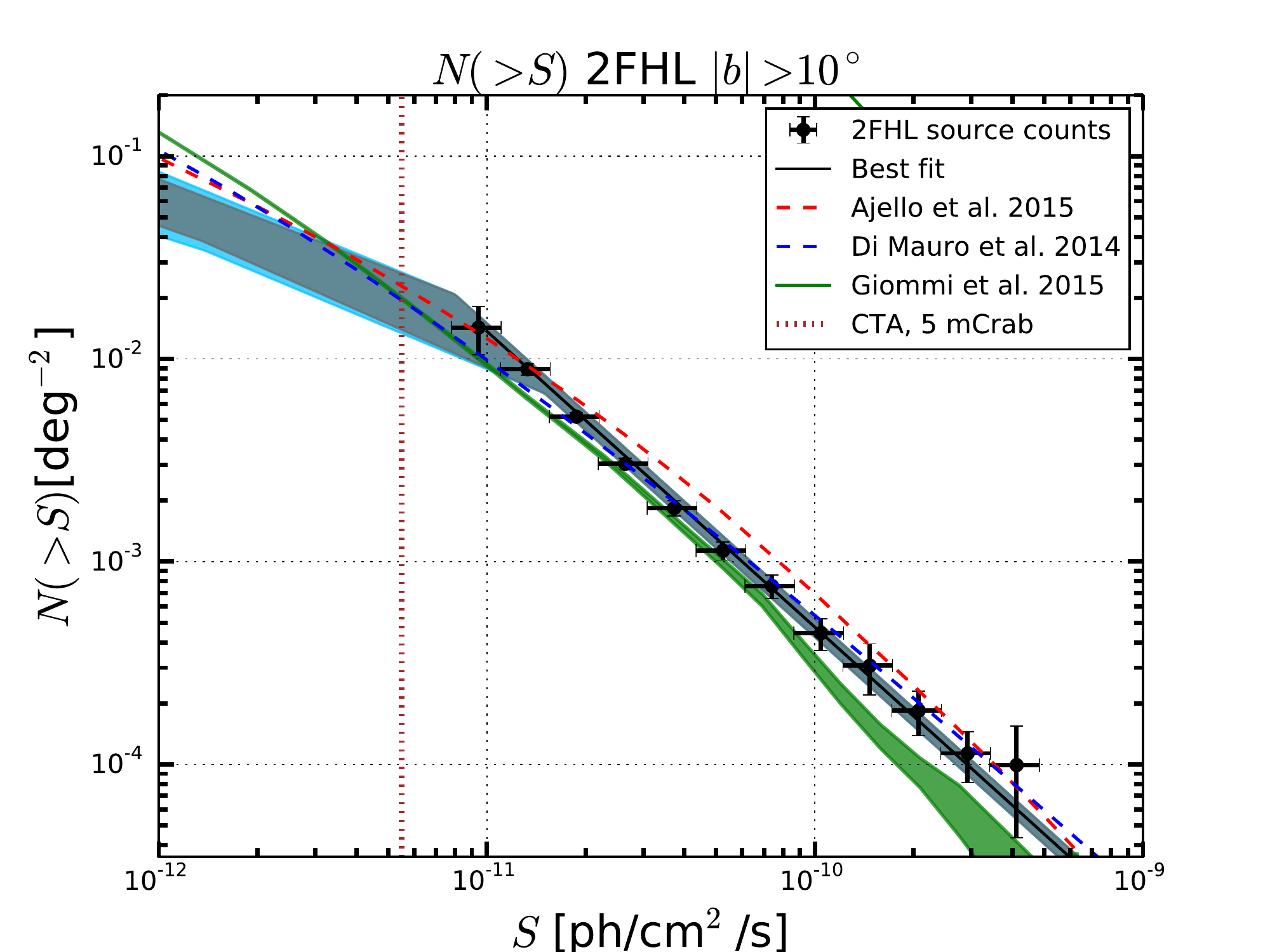}
\caption{Cumulative source count distribution $N(>S)$ with the uncertainty bands as in Fig.~\ref{fig:efficiency} together with the theoretical predictions from Ref.~\cite{DiMauro:2013zfa} (blue dashed line), \cite{Ajello:2015mfa} (red dashed line) and \cite{2015MNRAS.450.2404G} (green band). 
The vertical dotted brown line shows the 5\,mCrab flux reachable by CTA in 240 hrs of exposure \cite{2013APh....43..317D}.}
\label{fig:ncount} 
\end{figure}

The results from analyzing the sources in the simulated data can be used
to measure the detection efficiency $\omega(S)$, which is a weighting factor that takes into
account the probability to detect a source as a function of flux. 
%correcting for the   selection effects that the instrument, the background, and the detection pipeline may introduce.
The detection efficiency is simply derived from the simulations
measuring the ratio between the number of detected sources and the number
of simulated ones as a function of measured source flux.
%Once we have analyzed the 10 simulations with the prescriptions explained above, we derive the efficiency $\omega(F)$. We	divide the flux in a sampling of bins and for each bin we calculate the ratio between the number of detected sources and the number of sources in the simulations with a flux within this flux bin. After doing this procedure for each flux bin we derive the efficiency.
The result reported in Fig.~\ref{fig:efficiency} shows that the LAT detects any source in the $|b|>10^{\circ}$ sky for fluxes larger than $\approx 2 \times 10^{-11}$ ph cm$^{-2}$ s$^{-1}$, but misses 80--90\,\% of the sources with fluxes 
of $\approx 1 \times 10^{-11}$ ph cm$^{-2}$ s$^{-1}$ and many more below this flux.
The peak ($\omega(S)>$1) clearly visible at a flux of $\approx2 \times 10^{-11}$ ph cm$^{-2}$ s$^{-1}$ is due to the Eddington bias.

A reliable estimate of the detection efficiency is fundamental in order to correct the observed flux distribution of the 2FHL catalog and in turn to derive the intrinsic source count distribution, which is obtained as:
%\begin{equation}
%\frac{dN}{dS} = \frac{1}{\Omega \Delta S}\sum^{N_{\Delta S}}_{i=1} \frac{1}{\omega(f_i)}\ \ \ \ [{\rm cm^2\ s\  deg^{-2}}],
%\end{equation}
\begin{equation}
\frac{dN}{dS} \left( S_i \right) = \frac{1}{\Omega \Delta S_i} \frac{N_{i}}{\omega(S_i)}\ \ \ \ [{\rm cm^2\ s\  deg^{-2}}],
\end{equation}
where $\Omega$ is the solid angle of the $|b|>10^{\circ}$ sky,  $\Delta S_i$ is the width of the flux bin, $N_{i}$ is the number of sources in each flux bin and $S_i$ is the flux at the center of a given bin $i$. 
We verified through simulations that this method allows us to retrieve the correct source count distribution as long as the distribution used in the simulations is a faithful representation of the real one.
%Despite the overestimation in measured flux for faint sources due to Eddington bias, the method described above allows us to retrieve the correct source count distribution.
%Despite the fact that at low fluxes, because of the Eddington bias, the measured flux is a biased estimate of the true source flux, our tests with simulations show that the method described above allows us to retrieve the correct source count distribution.

This is found to be consistent, down to the sensitivity of the 2FHL catalog ($\approx 8 \times 10^{-12}$ ph cm$^{-2}$ s$^{-1}$), with a power-law function with slope $\alpha_1=2.49\pm0.12$ (see right panel of Fig.~\ref{fig:efficiency}).
This best-fit value is consistent with the Euclidean expectation
and motivated us to choose $\alpha_1 = 2.5$ in the simulations.
%The result is showed in the right panel of Fig.~\ref{fig:ncount} where the observed $dN/dS$ of the 2FHL and the intrinsic flux differential distribution are displayed.  The intrinsic $dN/dS$ follows the power-law trend given by $\propto S^{-\alpha_1}$ up to $7-8 \times 10^{-12}$ ph cm$^{-2}$ s$^{-1}$ where the correction given by the efficiency is about one order of magnitude. The uncertainties in the shape of the flux distribution below the threshold of the 2FHL catalog is also depicted in the right panel of Fig.~\ref{fig:efficiency}. As explained in the pixel counting part, this band is obtained considering the convolution of the choice for the break flux in the range $ S_b \in [8 \times 10^{-12},1.5 \times 10^{-11}]$ with the corresponding values for the slope below the break in the range $\alpha_2\in[1.60,1.75]$. 

Fig.~\ref{fig:ncount} shows the cumulative source count distribution that is defined as:
\begin{equation}
N(>S) = \int^{S_{\rm{max}}}_{S} \frac{dN}{dS'} \, dS'\ \ \ \ [{\rm deg^{-2}}],
\end{equation}
where $S_{\rm{max}}$ is fixed to be $10^{-8}$\,ph cm$^{-2}$ s$^{-1}$.% We point out that the results of our analysis are not affected by the choice of $S_{\rm{max}}$.}
%, is found to be compatible with a Euclidean distribution ($\propto S^{-2.5}$) and, implies a source density at those fluxes of about 0.015 deg$^{-2}$ at $8 \times 10^{-12}$\,ph cm$^{-2}$ s$^{-1}$.

%%%
%%% FLUCTUATION ANALYSIS
%%% 

In order to infer the shape of the $dN/dS$ below  the flux threshold for detecting point sources we have performed a photon fluctuation analysis. This helps us to probe the source count distribution to the level where sources contribute on average 0.5\,photons each.
The analysis is performed by comparing the histogram of the pixel counts of the real sky
with the ones obtained via Monte Carlo simulations and allows us to constrain
the slope of the differential flux distribution  below the threshold of the survey \cite{1993A&A...275....1H,2011ApJ...738..181M}.
We consider a differential flux distribution described as a broken power law where the slope above the break is $\alpha_1=2.5$ as determined in this work
while below the break the slope varies in different simulations between $\alpha_2\in[1.3, 2.7]$.
%We generate for each value of $\alpha_2$ 20 simulations. 
For each value of the slope we derive the model pixel count distribution averaging over the pixel count distributions obtained from 20 simulations.
%The simulated and real maps have been pixelized using the HEALPix tool \footnote{See http://healpix.sourceforge.net} \cite{2005ApJ...622..759G}. We have used a resolution of order 9, which translates into 3145728 pixels and an pixel size of about $0.11^{\circ}$, matching the size of the Pass 8 point-spread function. Consistent results are obtained when using a resolution of order 8. We consider a single energy bin from 50 GeV to 2 TeV.
The simulated and real maps have been pixelized using the HEALPix tool \footnote{See http://healpix.sourceforge.net} \cite{2005ApJ...622..759G}. We have used a resolution of order 9, which translates into 3145728 pixels and an pixel size of about $0.11^{\circ}$. Consistent results are obtained when using a resolution of order 8. We consider a single energy bin from 50 GeV to 2 TeV.

\begin{figure}
	\centering
	\includegraphics[width=1.1\columnwidth]{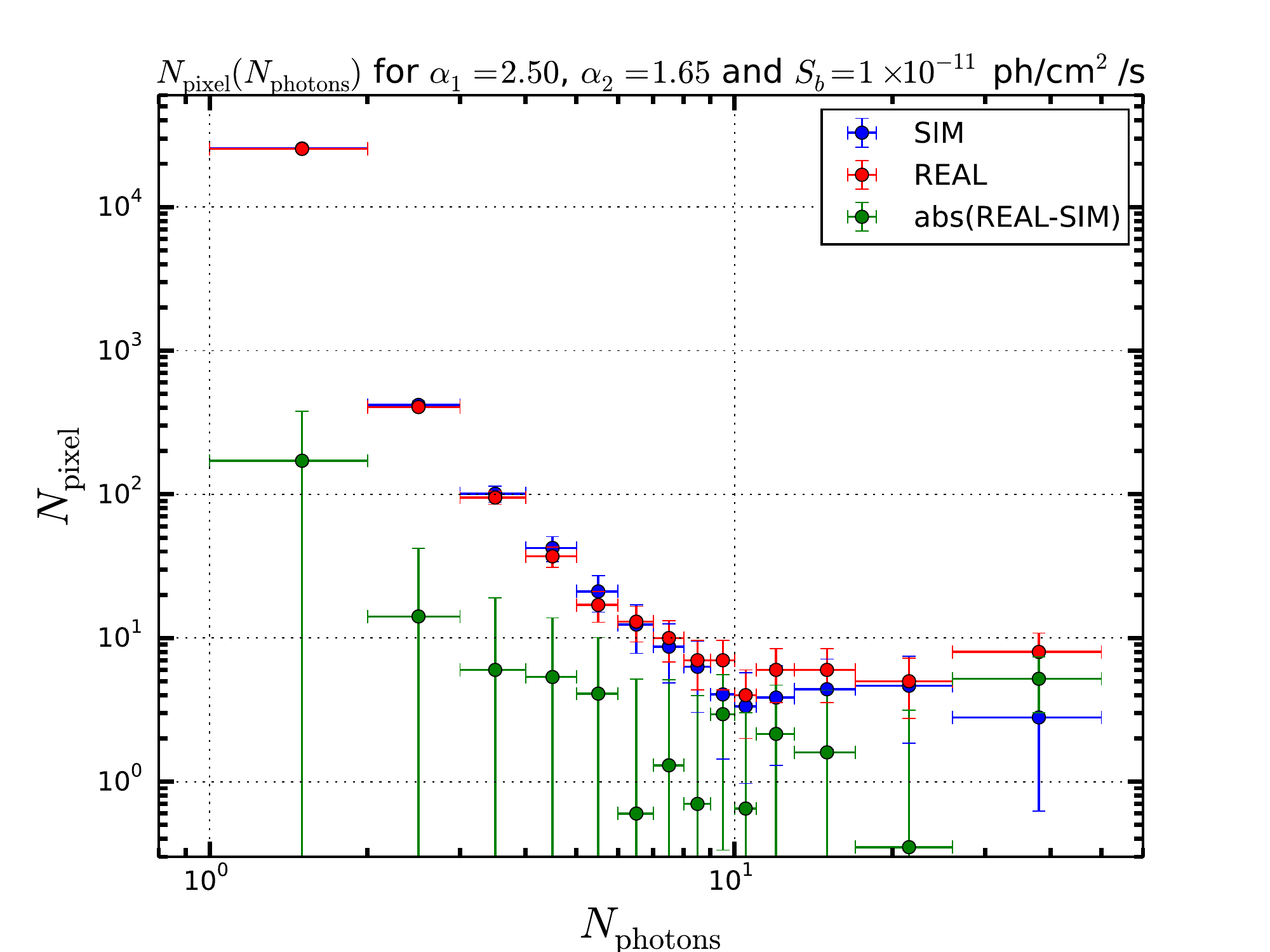}
\caption{Comparison between the pixel count distribution from the average of 20 simulations (blue points), and the distribution from the real sky (red points). The green points show the difference between the two distributions. In each number of photon bin $N_{\rm{photons}}$ ranging between $[N_{\rm{photon},1},N_{\rm{photon},2}]$ we display $N_{\rm{pixel}}$ with $N_{\rm{photons}}\in[N_{\rm{photon},1},N_{\rm{photon},2})$.}
\label{fig:pixel} 
\end{figure}

The model (averaged) pixel count distributions are compared to the real data using a $\chi^2$ analysis to determine the most likely scenario.
%Finally we performed a $\chi^2$ analysis comparing the average simulation and real sky map pixel counting distributions.
As expected, there is a degeneracy between the best-fit value of the slope $\alpha_2$ and the choice of the break flux, $S_b$. 
%{\bf The smaller the} value of the break flux, {\bf the larger} the break $|\alpha_1-\alpha_2|$ (i.e.{\bf ,} the smaller is the best{\bf -}fit value of the slope $\alpha_2$).  
%As expected{\bf ,} there is a degeneracy between the best{\bf -}fit value of the slope $\alpha_2$ and the choice of the \bf break flux, $S_b$. {\bf The smaller the} value of the break flux, {\bf the larger} the break $|\alpha_1-\alpha_2|$ (i.e.{\bf ,} the smaller is the best{\bf -}fit value of the slope $\alpha_2$).  
%Mattia has commented the following two sentences
%Considering the following values for the {\bf break flux} $S_b \in [0.6,0.8,1,1.5,2]\times 10^{-11}$ ph cm$^{-2}$ s$^{-1}$, the best{\bf -}fit values of the slope below the break are respectively $1.40 \pm 0.10$, $1.60\pm0.03$, $1.65\pm0.03$, $1.75 \pm 0.03$, $1.85\pm0.02$ while{\bf ,} the $\chi^2$ values are 25.0, 16.5, 12.4, 14.6 and 17.0 (for 12 degrees of freedom). 
%The best configuration {\bf has} a {\bf break} flux at $10^{-11}$ ph cm$^{-2}$ s$^{-1}$ and a slope $\alpha_2=1.65$. 
The result of the analysis is that the break flux is limited to the range between $S_b \in [8 \times 10^{-12},1.5 \times 10^{-11}]$\,ph cm$^{-2}$ s$^{-1}$ while the index below the break is in the range $\alpha_2\in[1.60,1.75]$.
The best configuration, which we refer to as our benchmark model, has a break flux at $1\times 10^{-11}$ ph cm$^{-2}$ s$^{-1}$ and a slope $\alpha_2=1.65$ with a $\chi^2=12.4$ (for 12 degrees of freedom). 
This implies that the source count distribution must display a hard break $|\alpha_1-\alpha_2|\approx0.9$ from the Euclidean behavior measured at bright fluxes.
We show in Fig.~\ref{fig:pixel}, for the best-fit configuration, the comparison between the pixel count distribution evaluated for the average of 20 simulations, and the same quantity as derived from the real data. The figure also shows the differences between these two distributions.
%Mattia has commented the following three sentences
%Indeed the $\chi^2$ value for a break flux lower than $0.8 \times 10^{-11}$ is much larger than those obtained for larger choices of the break flux. 
%This {\bf strongly disfavors} the scenario in which the source count distribution remains Euclidean down to fluxes of  $0.6 \times 10^{-11}$\,ph cm$^{-2}$ s$^{-1}$.
%On the other hand, the direct determination of the source count distribution of the 2FHL showed that it is compatible with a Euclidean distribution 
% does not contain any deviation from the slope $\propto S^{-\alpha_1}$ for fluxes $\gtrsim 1.5 \times 10^{-11}$ ph cm$^{-2}$ s$^{-1}$ (see right panel of Fig.~\ref{fig:efficiency}). 
%Therefore the fluctuation analysis limits the slope below the {\rm break flux} to vary in the range $\alpha_2\in[1.60,1.75]$. 
%Given the result of the pixel counting analysis, we have fixed the shape of the differential flux distribution to be a broken power law with the break at $10^{-11}$ ph cm$^{-2}$ s$^{-1}$ and a slope above/below the break of $\alpha_1=2.5$ and $\alpha_2=1.65$.

The lowest flux that the photon fluctuation analysis is sensitive to can be estimated by adding to the source count distribution one more break flux below that of the benchmark model. 
%We fixed the slope below this second break equal to $\alpha_3=$1.80. This model is already rejected by our analysis (see above) and the break flux is varied in the range $S_{\rm{lim}}\in[5 \times 10^{-13},5 \times 10^{-12}]$ ph cm$^{-2}$ s$^{-1}$ to register when a worsening of the $\chi^2$ (with respect to the best-fit one) is observed.
We fixed the slope below this second break to $\alpha_3=$1.80, which is at the edge of the derived range for $\alpha_2$, while the break flux is varied in the range $S_{\rm{lim}}\in[5 \times 10^{-13},5 \times 10^{-12}]$ ph cm$^{-2}$ s$^{-1}$ to register when a worsening of the $\chi^2$ (with respect to the best-fit one) is observed.
%We already know that this value is inconsistent with the value of our benchmark model which is 1.65 below the break fixed to $10^{-11}$ ph cm$^{-2}$ s$^{-1}$. However we move the position of the second break in the range $S_b \in [5 \times 10^{-13},5 \times 10^{-12}]$ ph cm$^{-2}$ s$^{-1}$ and when the second break flux value is below the sensitivity of the pixel counting method, i.e. for flux values which generate on average 0 photons, the $\chi^2$ value, for the comparison of the simulation and real sky pixel counting distribution, would be the same as in the benchmark model with $\alpha_1=1.65$.  On the other hand when the position of the second break will be inside the sensitivity range of the method the incorrect slope of $1.80$ would worsen the value of the $\chi^2$.
The result of this analysis is that the fit worsened by more than 3\,$\sigma$ for $S_{\rm{lim}}\gtrsim 1.3 \times 10^{-12}$\,ph cm$^{-2}$ s$^{-1}$.
%, fixing the sensitivity $S_{\rm{lim}}$ to the position of the second break which results in a deviation of $3\sigma$ from the best fit $\chi^2$ value of the benchmark model, $S_{\rm{lim}}\approx 1.3 \times 10^{-12}$ ph cm$^{-2}$ s$^{-1}$.
%The pixel counting methods therefore permits us to infer the shape of the LogN-LogS up to a factor of about 6 below the threshold of the 2FHL catalog.
The results of the photon fluctuation analysis are reported in Figs.~\ref{fig:efficiency} and \ref{fig:ncount}, which show that this technique allows us to measure the source count distribution over almost three decades in flux.

We have tested also the possibility that a new source population could emerge in the flux distribution with a Euclidean distribution, as might be expected, for example, from star-forming galaxies \cite{2012ApJ...757L..23B}. In this test we set  $\alpha_3=2.50$ and follow the method described above to derive the maximum flux at which a possible re-steepening of the source counts might occur. 
%This is found to be $S_{\rm{lim}}\approx 7 \times 10^{-13}$\,ph cm$^{-2}$ s$^{-1}$ and Fig.~\ref{fig:efficiency} (right panel) shows that the integrated emission of such a population would quickly exceed (at fluxes of $\sim2\times10^{-13}$ ph cm$^{-2}$ s$^{-1}$) the totality of the EGB intensity.
This is found to be $S_{\rm{lim}}\approx 7 \times 10^{-13}$\,ph cm$^{-2}$ s$^{-1}$ and the integrated emission of such a population would exceed at fluxes of $\sim7\times10^{-14}$ ph cm$^{-2}$ s$^{-1}$ the totality of the EGB intensity.

%We have therefore added (similaras done above a second break but at this time with $\alpha_3=$2.50. Following the same method employed before we infer an upper limit for the second break equal to $S_{\rm{lim}}\approx 7 \times 10^{-13}$ ph cm$^{-2}$ s$^{-1}$. Strong constraints on such population are also derived using the maximal contribution point sources can produce on the EGB. In the right panel of Fig.~\ref{fig:efficiency} for example a re-steepening of the LogN-LogS would overestimate the contribution to EGB at $2\times10^{-13}$ ph/cm$^2$/s}

Our best-fit model for the flux distribution $dN/dS$ is therefore, for $S \gtrsim10^{-12}$\,ph cm$^{-2}$ s$^{-1}$, a broken power-law with break flux in the range $S_b \in [0.8,1.5] \times 10^{-11}$, slopes above and below the break of $\alpha_1=2.49\pm0.12$ and $\alpha_2\in[1.60,1.75]$, respectively and a normalization $K=(4.60 \pm 0.35)\times 10^{-19}$ deg$^{-2}$ ph$^{-1}$ cm$^{2}$ s. We believe this describes the source counts of a single population (blazars), because no re-steepening of the source count distribution is observed and because the large majority (97\,\%) of the detected sources are likely blazars.

Fig.~\ref{fig:ncount} reports the theoretical expectations for the source count distribution given by blazars \cite{2015MNRAS.450.2404G,Ajello:2015mfa} and BL Lacs \cite{DiMauro:2013zfa}.
These models are consistent with the observations at bright fluxes, but are above the experimental $N(>S)$ by about a factor of 2 at $S = 10^{-12}$ ph cm$^{-2}$ s$^{-1}$.
We include in the same figure also the predicted 5\,mCrab sensitivity reachable by CTA in 240 hours  in the most sensitive pointing strategy \cite{2013APh....43..317D}. At these fluxes the source density is $0.0194 \pm 0.0044$ deg$^{-2}$, which translates to the serendipitous detection of $200\pm45$ blazars in one quarter of the full sky. It is also interesting to note that our analysis constrains the source count distribution to fluxes that are much fainter than those reachable by CTA in short exposures. 
%We make note that the threshold of the 2FHL is about $8-9 \times 10^{-12}$ ph cm$^{-2}$ s$^{-1}$ while the CTA sensitivity, in this detection strategy, which is the most promising one, is about $5-6 \times 10^{-12}$ ph cm$^{-2}$ s$^{-1}$. The {\it Fermi}-LAT experiment has therefore already resolved almost all the $\gamma$-ray sky CTA will observe.

%PIXEL COUNTING

%%% CONCLUSIONS
Once known, the source count distribution can be used to estimate the contribution of point sources to the EGB. This is performed by integrating the flux distribution $dN/dS$ as follows:
\begin{equation}
\label{eq:igrb}
I =  \int_{0}^{S_{\rm{max}}} \, S' \, \frac{dN}{dS'} dS'\ \ \ \ [{\rm ph\,cm^{-2}\,s^{-1}\,sr^{-1}}].
\end{equation}
Choosing ${S_{\rm{max}}}=10^{-8}$ \,ph cm$^{-2}$ s$^{-1}$ we find that the total integrated flux from point sources is $2.07^{+0.40}_{-0.34}\times 10^{-9}$\,ph cm$^{-2}$ s$^{-1}$ sr$^{-1}$ which constitutes $86^{+16}_{-14} \%$ of the EGB above 50\,GeV estimated in \cite{Ackermann:2014usa}. This validates the predictions of models \cite{DiMauro:2013zfa,DiMauro:2015tfa,Ajello:2015mfa}. 
Point sources with fluxes $S>1.3\times 10^{-12}$\,ph cm$^{-2}$ s$^{-1}$ produce $1.47^{+0.20}_{-0.24}
\times 10^{-9}$\,ph cm$^{-2}$ s$^{-1}$ sr$^{-1}$, while 
$6.0^{+2.0}_{-1.0}\times 10^{-10}$\,ph cm$^{-2}$ s$^{-1}$ sr$^{-1}$ is produced by sources below
that flux.

The {\it Fermi}-LAT has measured the angular power spectrum of the diffuse $\gamma$-ray background at $|b|>30^{\circ}$ and in four energy bins spanning the 1-50\,GeV energy range \cite{2012PhRvD..85h3007A}.
For multipoles $l\geq155$ the angular power $C_P$ is found to be almost constant, suggesting that the anisotropy is produced by an unclustered population of unresolved point sources.
Indeed, Refs.~\cite{Cuoco:2012yf,DiMauro:2014wha,Chang:2013ada} argue that  most of the angular power measured by the {\it Fermi}-LAT is due to unresolved emission of radio-loud active galactic nuclei.

%Since we have derived the logN-logS for energies larger than 50 GeV we can predict the angular power in this energy range in the following way:
The angular power due to unresolved sources at $>$50\,GeV can be readily predicted from the source count distribution as:
\begin{equation}
C_P =  \int_0^{S_{\rm{max}}} \left( 1-\omega(S') \right)S'^2 \frac{dN}{dS'} dS'  [{\rm (ph\ cm^{-2}\ s^{-1})}^2 {\rm sr}^{-1} ],
\end{equation}
%{\bf I would skip this formula writing only. Since we have derived the logN-logS for energies larger than 50 GeV we can predict the angular power in this energy range using the same method as in Eq.~\ref{eq:igrb} but with  $\left( 1-\omega(S) \right)S^2 \frac{dN}{dS} $ as integrand.}
The angular power evaluates to $C_{P}(E>50\,{\rm GeV}) = 9.4^{+1.0}_{-1.6} \times 10^{-22}$ (ph/cm$^2$/s)$^2$  sr$^{-1}$. This is the first observationally-based prediction of the angular power at $>$50\,GeV.
%Our estimation for $C_{P}(E>50\,{\rm GeV})$ is in good agreement with the extrapolation of the {\it Fermi}-LAT angular power measurements.
Our estimation for $C_P (E > 50 GeV)$ is in good agreement with the extrapolation of the {\it Fermi}-LAT angular power measurements \cite{2012PhRvD..85h3007A} above 50\,GeV and is consistent with the calculated anisotropy due to radio loud active galactic nuclei made in Refs.~\cite{Cuoco:2012yf,DiMauro:2014wha}.
%Indeed the fit to {\it Fermi}-LAT data, with a power law, at 80 GeV has the same value of $C_{P}$ we derive for $>50$ GeV.
%Moreover the value of $C_{P}(E>50\,{\rm GeV})$ is consistent with the calculated anisotropy due to radio loud active galactic nuclei made in \cite{DiMauro:2014wha}.

%Mattia has commented the following part.
%Ref.~\cite{broderick2012} introduced an alternative model for the luminosity function and evolution of spectrally hard blazars. In that model, any cascade component generated in the interaction of TeV photons with the extragalactic background light is suppressed due to plasma beam instabilities. 
%The prediction of the angular power, provided by  \cite{broderick2014} up to
%30\,GeV, if extrapolated at $>$50\,GeV with a power law  is found in tension with our measurement.
%However, the extrapolation of the angular power derived with this model up to about 30 GeV \cite{broderick2014} is in strong tension with our estimate at $E>50$ GeV.
%{\bf The predicted angular power provided by \cite{broderick2014} up to 30 GeV, if extrapolated to $>$50 GeV with a power law, is found to be in tension with our measurement.}

In conclusion, the {\it Fermi}-LAT collaboration has used the new event-level analysis Pass~8 to conduct an all-sky survey above 50 GeV. The resulting 2FHL catalog contains 253 sources at $|b|>10^{\circ}$ and closes the energy gap between the LAT and Cherenkov telescopes.
We have thoroughly studied the properties of both resolved and unresolved sources in the 50\,GeV--2\,TeV band using detailed Monte Carlo simulations and a photon fluctuation analysis. This allowed us to characterize, for the first time, the source count distribution above 50\,GeV, which is found to be compatible at $\gtrsim10^{-12}$\,ph cm$^{-2}$ s$^{-1}$ with a broken power-law model with a break flux in the range $S_b \in [0.8,1.5] \times 10^{-11}$\,ph cm$^{-2}$ s$^{-1}$, and slopes above and below the break of, respectively, $\alpha_1=2.49\pm0.12$ and $\alpha_2\in[1.60,1.75]$.
% {\bf and a normalization $K=(4.60 \pm 0.35)\times 10^{-19}$ deg$^{-2}$ ph$^{-1}$ cm$^{2}$ s.
A photon fluctuation analysis constrains a possible re-steepening of the flux distribution to a Euclidean behavior ($\alpha_3=2.50$) to occur at fluxes lower than $\sim 7\times10^{-13}$\,ph cm$^{-2}$ s$^{-1}$.
%This allowed us to estimate that blazars, and in particular BL Lacs, explain almost the totality (96$^{+15}_{-18}$\,\%) of the $>$50\,GeV IGRB 
Our analysis permits us to estimate that point sources, and in particular blazars, explain almost the totality (86$^{+16}_{-14}$\,\%) of the $>$50\,GeV EGB.

This might have a number of important consequences, since any other contribution, exotic or not, must necessarily be small.
%Using the photon fluctuation analysis to infer the shape of $dN/dS$ below the flux threshold, we derive that the differential flux distribution is given by a broken power law with a break flux in the range $S_b \in [0.8,1.5] \times 10^{-11}$ and a slope above and below the break of $\alpha_1=2.49\pm0.12$ and $\alpha_2\in[1.60,1.75]$. Considering this flux distribution we create simulations of the $\gamma$-ray sky at energy $E>50$ GeV and we analyze them resulting on average in 270 detectable sources at $|b|>10^{\circ}$. This analysis permits us to derive the efficiency of the 2FHL catalog $\omega(S)$ and to infer the real source count distribution of the catalog. We predict that point sources are able to explain between $80-100\%$ of the IGRB at $E>50$ GeV.
%The results presented in this paper has an important impact on different field of astroparticle physics.
%Since we have derived that the $96\%^{+15}_{-18}$ of the IGRB at $E>50$ GeV is explained with extragalactic point sources, the room left to other contributions is constrained to have a $1\sigma$ upper limit equal to $19\%$ of the IGRB in the same energy range.
This bound might imply strong constraints for the annihilation cross section or decay time of high-mass dark matter particles producing $\gamma$-rays \cite{DiMauro:2015tfa,Ajello:2015mfa}.
Tight constraints could also be inferred on other $\gamma$-ray emission mechanisms due to other diffusive processes such as UHECRs \cite{2011PhRvD..84h5019A,2012JCAP...01..044G}. Finally, if the neutrinos detected by IceCube have been generated in hadronic cosmic-ray interactions, then the same sources producing the neutrino background will produce part of the sub-TeV $\gamma$-ray background \cite{Aartsen:2014gkd}.
Because blazars were found not to be responsible for the majority of the neutrino flux \citep{gluesenkamp2015}, the fact that the 50\,GeV--2\,TeV $\gamma$-ray background is almost all due to blazars constrains the contribution of other source classes to the neutrino background. Such constraints will be presented in a dedicated paper.
%Our results can have also interesting consequences for the study of the origin of Ice Cube neutrino flux \cite{Aartsen:2014gkd}. This topic will be addressed in a dedicated paper.

%All these topics are related to the search for dark matter \cite{Ackermann:2015tah,DiMauro:2015tfa}, the existence of plasma beam instabilities that my suppress the cascade components of blazars \citep{broderick2012}, contribution to the high energy $\gamma$-ray sky of diffuse processes such as the interaction of UHECRs with the extragalactic background light \cite{2011PhRvD..84h5019A,2012JCAP...01..044G} and the interpretation of the high-energy astrophysical neutrino flux measured by IceCube \cite{Aartsen:2014gkd}.%
%{\bf ho aggiunto anche l'argomento di Ice Cube e dei meccanismi diffusivi esotici}.

\begin{acknowledgements}
The \textit{Fermi}-LAT Collaboration acknowledges support for LAT development, operation and data analysis from NASA and DOE (United States), CEA/Irfu and IN2P3/CNRS (France), ASI and INFN (Italy), MEXT, KEK, and JAXA (Japan), and the K.A.~Wallenberg Foundation, the Swedish Research Council and the National Space Board (Sweden). Science analysis support in the operations phase from INAF (Italy) and CNES (France) is also gratefully acknowledged.
\end{acknowledgements}

\bibliography{2fhl}

\end{document}